\documentclass[aps,10pt,twocolumn,showpacs]{revtex4}
\usepackage{amssymb}
\usepackage{graphicx,setspace}
\usepackage{amsmath}

\begin{document}
\title{The flow induced by a mock whale : origin of flukeprint formation.}
\author{Germain Rousseaux.}
\affiliation{Universit\'{e} de Nice-Sophia Antipolis,
Laboratoire J.-A. Dieudonn\'{e}, UMR CNRS-UNS 6621,\\
Parc Valrose, 06108 Nice Cedex 02, France, European Union.}


\begin{abstract}

Fluke oscillations of cetaceans  (whales, dolphins and porpoises) imprint the ocean surface with an oval patch that surface waves cannot enter.  Scientists have hypothesized that modification of surface tension by surfactants creates flukeprints.  Here, we show, on the contrary, that the formation of flukeprint is primarily due to a wave-current interaction problem.  We provide Particle Image Velocimetry measurements of anisotropy and vorticity in the flow generated by a mock whale fluke in a laboratory experiment. We explain for the first time why long gravity waves enter the flukeprint, whereas short gravity waves are blocked. 

\end{abstract}


\maketitle

\begin{figure}[htbp!]
\includegraphics[scale = .115]{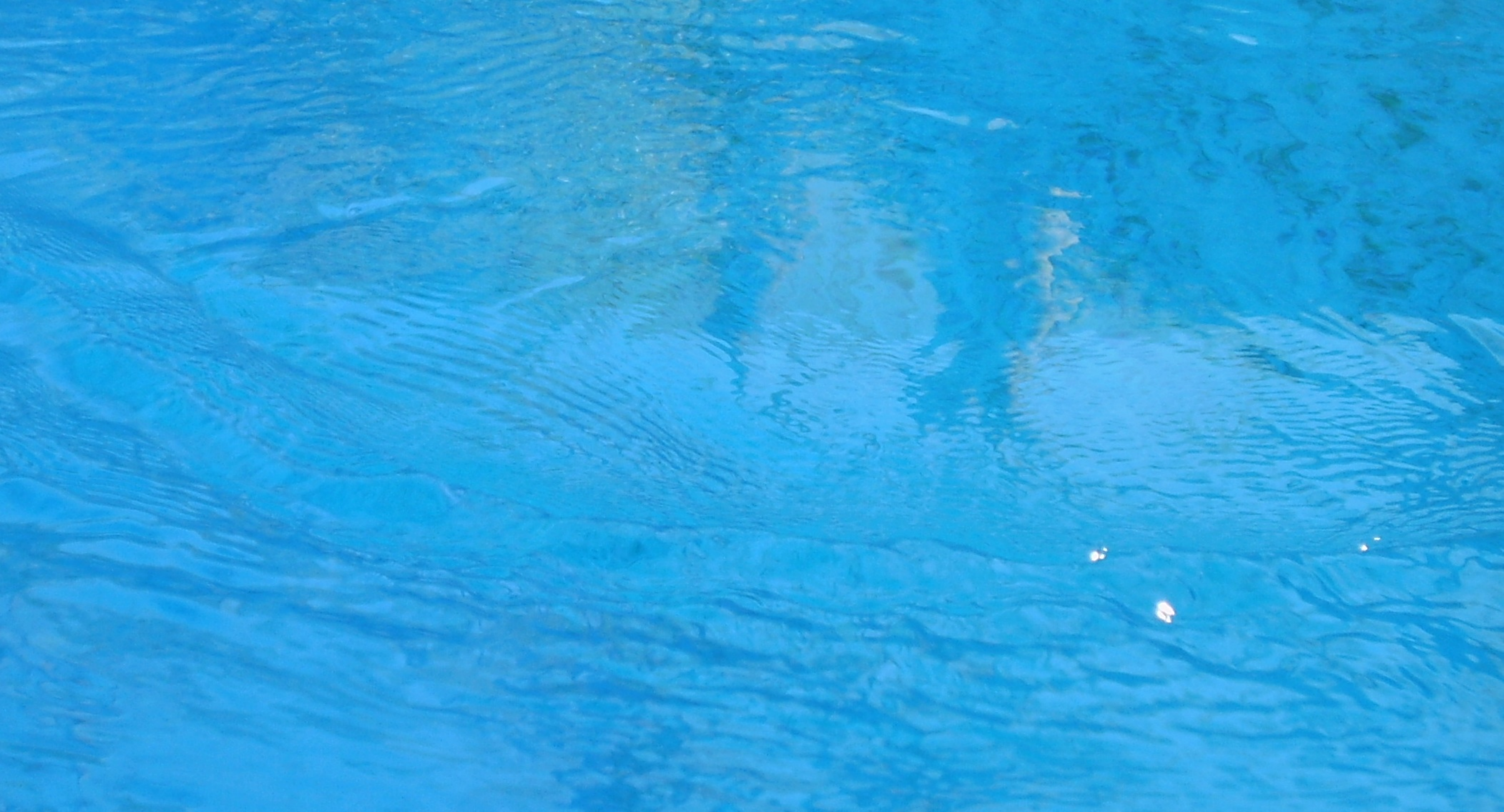}
\includegraphics[scale = .074]{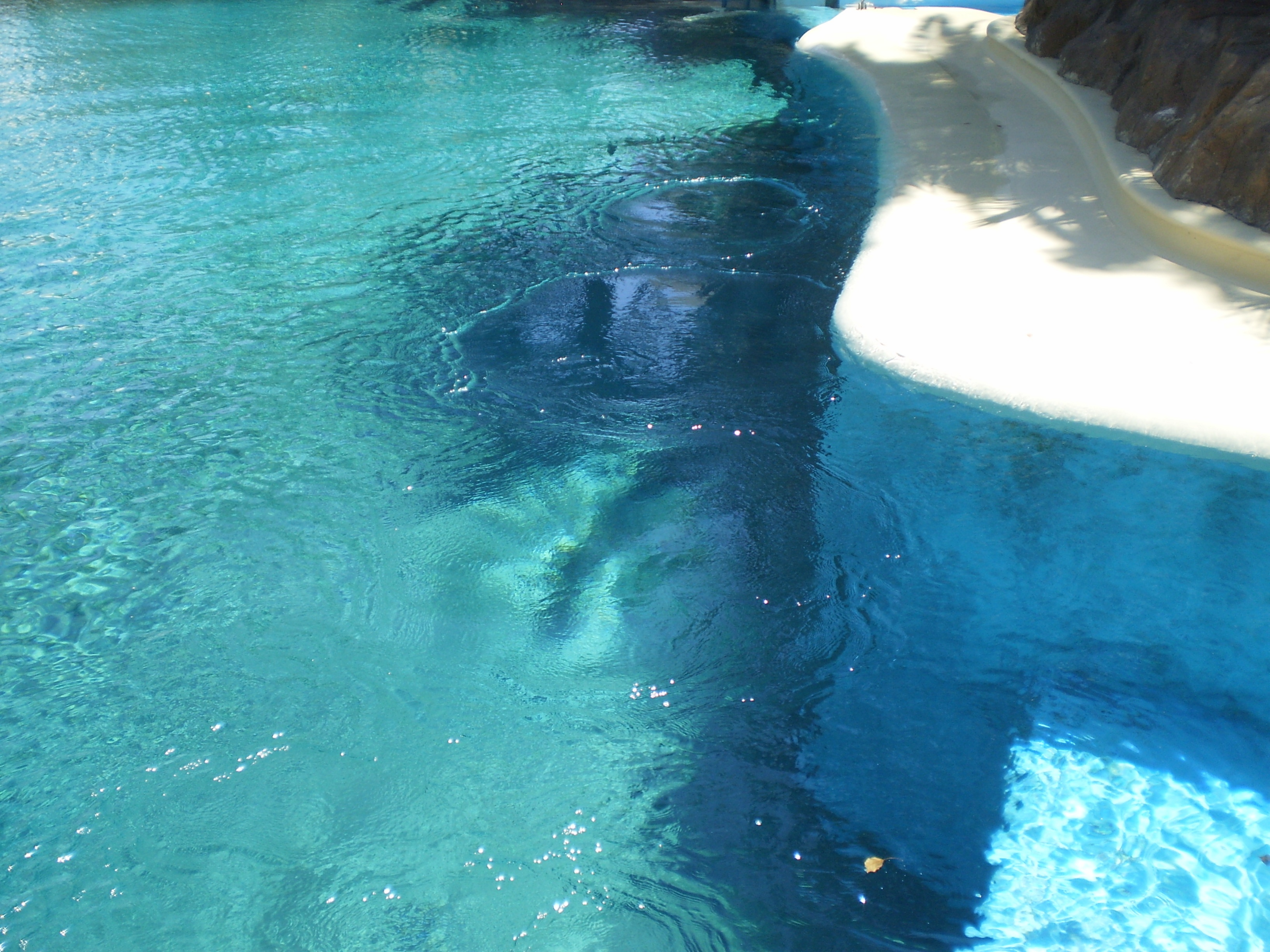}
\caption{Dolphin's flukeprints. Top: A close-up of the edge of a dolphin flukeprint where mode-converted capillary waves can be observed near the boundary and are subsequently dampened in the print. Bottom: A sequence of four flukeprints created by a dolphin.  The flukeprints grow in size over time, so the small prints are the ones made most recently by the dolphin traveling from "bottom to top" (dolphin fluke is visible at the top of the picture).}
\label{fig:dolphin}
\end{figure}

Free propagation of waves is an exception. For example, in a marine environment, water waves are refracted by the varying landscape of the ocean floor as they approach the seashore and ocean currents modify wave propagation. We consider wave interactions that have fascinated both casual whale-watchers and marine biologists: cetacean flukeprints.  Cetacean flukeprints (made by whales, dolphins and porpoises) occur when a vortex ring, shed by the fluke, interacts with waves present on the free surface of the water \cite{Levy:2011}.  In the ocean these patterns are sometimes known as  ``whale footprints''.  The surface signatures are characterized by a smooth oval pattern where short gravity waves cannot penetrate.  Depending on flow conditions, one can also observe several radii surrounding the smooth oval corresponding to mode conversion into blue-shifted and capillary waves as can be seen for dolphins flukeprints easily observed in a swimming pool (Fig. \ref{fig:dolphin}).  The key explanation of flukeprint formation, first revealed in the present work, involves the dispersive properties of water waves along with the effect of gravity, as well as surface tension and wave-breaking in the mode conversion induced by the surface current.

Similar surface patterns occur when a vortex ring created by an upward-pointing jet interacts with a free surface, although these patterns are simpler due to the regular shape of the jet orifice.  Another related example is a circular hydraulic jump in which a jet of water impacts a solid plate and creates a disk-shaped region with a boundary that surface waves cannot penetrate \cite{PRE11}. Long shallow water waves are blocked by the flow induced on the plate by the impacting jet with a characteristic border, namely the jump. In essence, the flukeprint is an inverted circular jump caused by a jet-like flow induced by a vortex ring obliquely impacting the ocean surface with different dispersive properties.

Here, we provide for the first time, PIV measurements from experiments reported in \cite{Levy:2011} where an oscillating 2D polypropylene fluke created vortex rings and resulting surface prints were visualized by particles and a laser sheet.

The landscape of waves on the ocean surface is complex, with interactions of both gravity
and capillary waves \cite{Mei}.  As argued in \cite{Levy:2011}, the current forming a flukeprint
originates below the surface at the swimming depth of the whale.  As a vortex ring forms, the current flows upwards through the center and then radially outwards at the ocean's surface, interacting with
oncoming gravity and capillary waves already present.

In deep water and without surface tension, when a gravity wave meets a counter-current,
the incident wavelength diminishes (the phenomenon known as blue-shifting) because of
normal dispersion.  That is, the wave group velocity is a decreasing function of
the wavenumber $k$. In this regime ($kh>>1$ where h is the water depth), the blocking of pure gravity waves (neglecting surface tension $\gamma =0$) is then described by the following formula \cite{PRL09}:
\begin{equation}
U_g=-\frac{gT}{8\pi}.
\label{Ug}
\end{equation}
where $T$ is the period of the incoming waves in the gravity field $g$. This equation indicates that the blocking velocity is linearly related to the wave period.  We expect wave blocking if the period of the existing gravity waves is short, but no blocking if the period is long.  {\em This simple formula explains the presence of only long waves inside the flukeprints and has not previously been noted}.  Since the blocking velocity in deep water depends on the wave period, it is an intrinsically dispersive effect whereas in shallow water the blocking occurs for a velocity which depends only on the water depth $h$ that is $U_h=\sqrt{gh}$ \cite{Peregrine, Dingemans}.

When including surface tension ($\gamma \neq 0$) in the gravity dispersion relation, Badulin et al. \cite{Badulin} observed that as long as the counter-current is strong enough, gravity waves can still be blocked in deep waters. In addition, blue-shifted waves appear by mode conversion but are also stopped while drifting backward at a new blocking boundary, formed where the blue-shifted wave merges with a new capillary solution \cite{Badulin, TM, Trulsen, NJP10}.  Capillary waves appear at this secondary blocking line, described in \cite{NJP10}, and propagate in the same direction as the incident gravity waves, provided the regime stays linear. Ultimately, the capillary waves no longer propagate far inside the flukeprint because they are not only quickly damped by viscosity but they are no longer solutions of the dispersion relation \cite{NJP10}, and are seen on its boundary. The threshold velocity for the appearance of capillary waves in deep water on a flow current is $U_\gamma-2\pi\frac{l_c}{T}$ where $U_\gamma =\sqrt{2}\left(\frac{\gamma g}{\rho}\right)^{1/4}$ is the minimum of the phase velocity as a function of the wavenumber and where $l_c=\sqrt{\frac{\gamma}{\rho g}}$ is the capillary length \cite{NJP10}. If the amplitude grows too much, wave breaking will lessen the conversion of the incident gravity waves to capillary waves.  In this case, energy is dissipated in turbulence rather than transferred linearly towards the capillary waves.

In addition to its influence on the mode conversion, surface tension can play another role in the flukeprint formation since it also changes because of variation in surfactant concentration or temperature. This phenomenon in which surface tension is non-increasing as the temperature or surfactant concentration increases is referred to generically as a Marangoni effect.  Oily skin sloughed by cetaceans may also modify the properties of the air-water interface, but in a different way than just changing the value of
surface tension. An oil monolayer can feature a surfactant concentration gradient (Gibbs-Marangoni effect) because of the motion of water waves \cite{calm}. This induces a surface dilatational elasticity (Gibbs surface elasticity) related to the presence of elastic longitudinal waves within the monolayer. The resulting rigidity of the surface dampens the capillary waves (which constitutes the Franklin effect) by
increasing the shear forces on the free surface hence the viscous dissipation.  Moreover,
capillary waves are naturally dampened by viscosity and their amplitude is known to
decreases exponentially with a factor proportional to the water viscosity $\mu$ and
inversely proportional to surface tension $\gamma$ and wave period $T$:
$A=A_0e^{-\frac{2\pi \mu }{T\gamma}}$ \cite{calm}. Hence, a decrease of surface tension quickly
dampens capillary waves. In addition to the blocking of short gravity waves by the induced flow created by the vortex ring, the Franklin effect may provide another mechanism for the damping of
capillary waves in addition to their usual dissipation by viscosity.

Recently, it was demonstrated \cite{Churnside} using thermal imagery that the inside of flukeprint has
a colder water temperature than the outside.  This temperature signature is a natural consequence of the vortex ring inducing a jet of water taking cold water from below the free surface and dispatching it at the surface forming the interior of the flukeprint. The thermal signature provides one way to track whales. However, it must be noted that the thermal change is not the origin of the print.  A decrease in temperature will increase the surface tension, hence decrease the viscous damping of the capillary waves.

\begin{figure}[!htbp!]
\includegraphics[scale = .23]{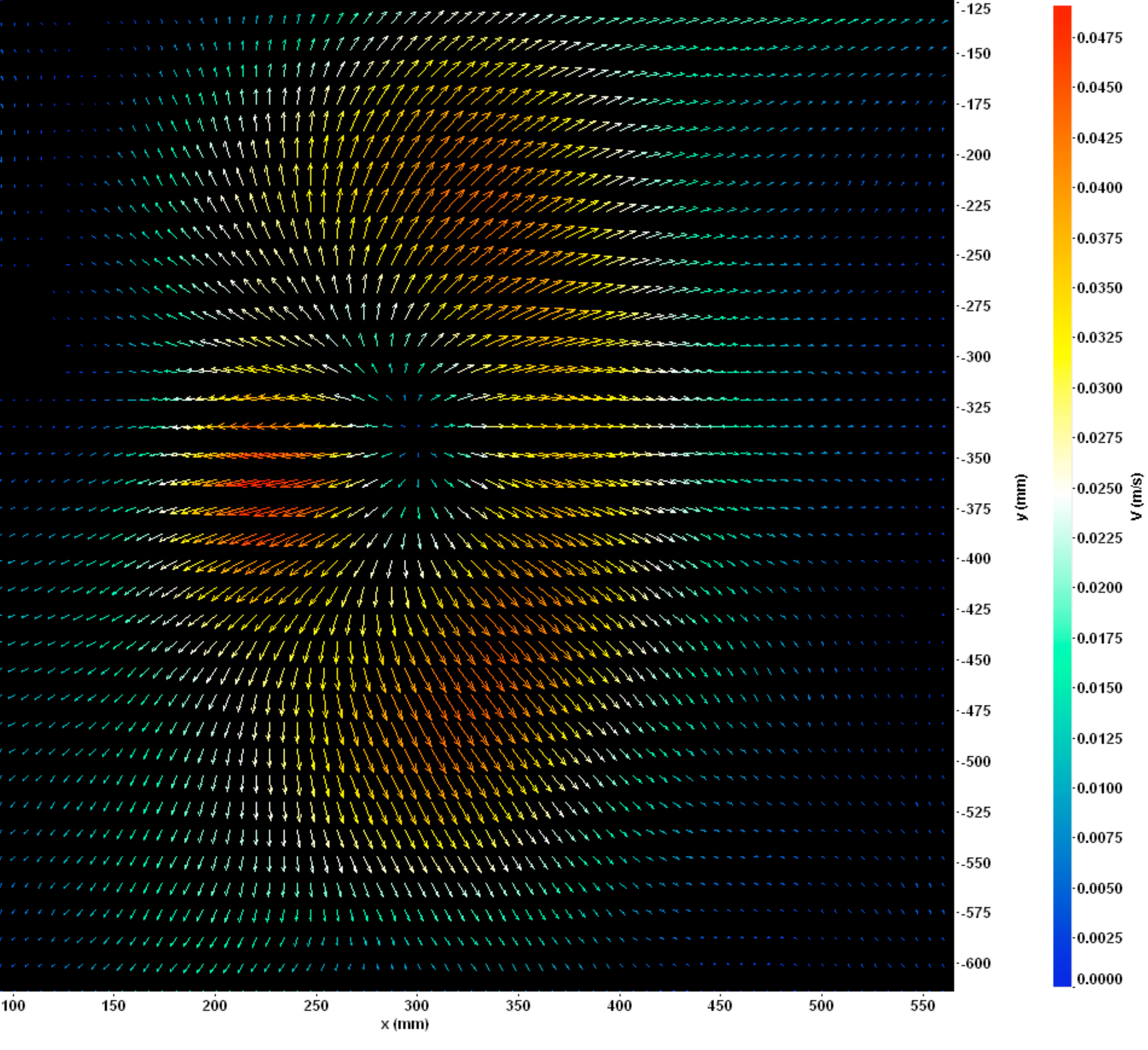}
\includegraphics[scale = .23]{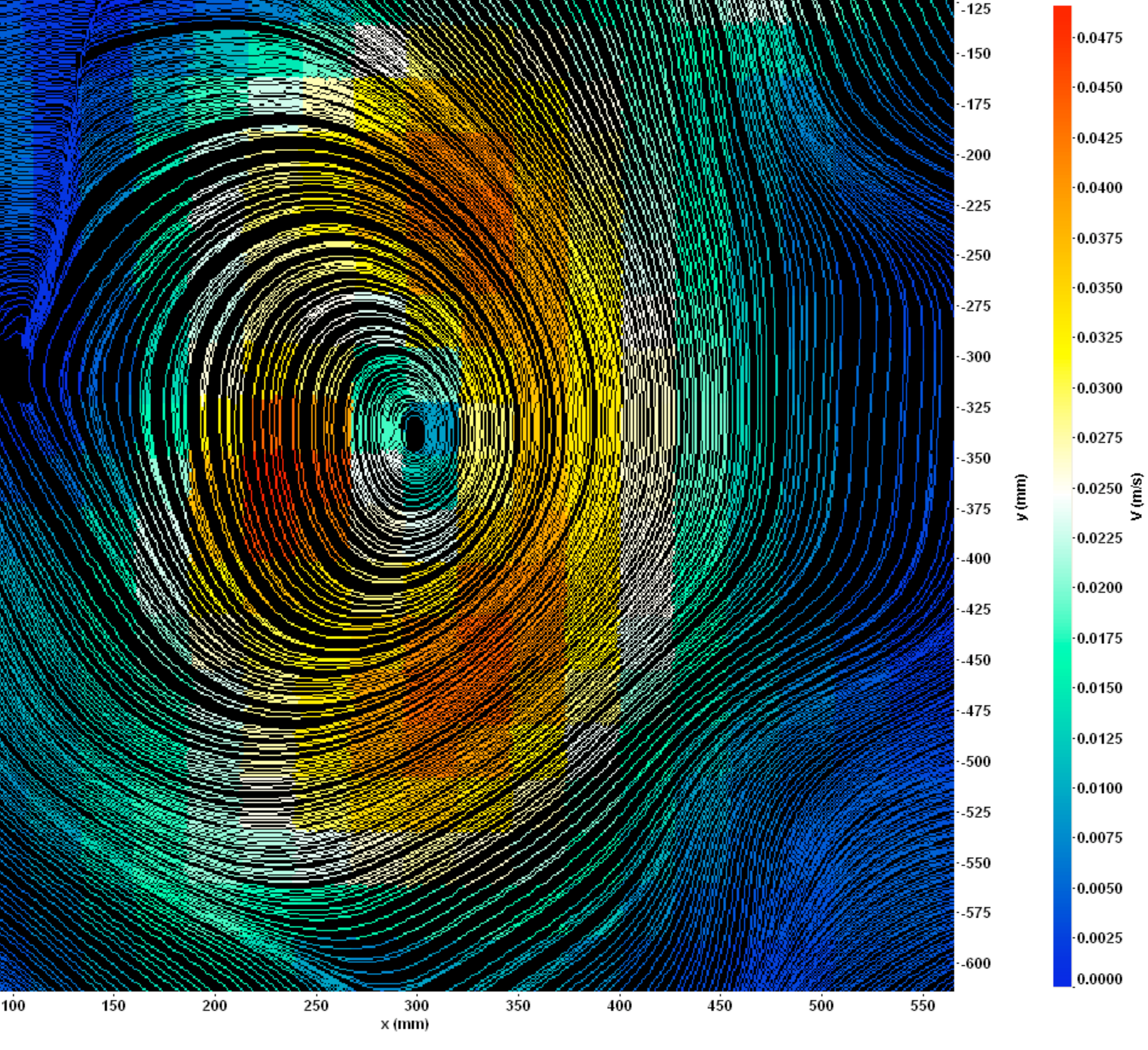}
\caption{Surface flows. Top: Typical instantaneous velocity field on the free surface created by an impacting vortex ring (t=0.266s after flapping begins). The fluke is oriented as if the whale is facing left and the flow is stronger on average on the right. The colors correspond to the modulus of the velocity at each point. The angle of the fluke at the maximum is $60^\circ$ and it oscillates through
a $25^\circ$ arc. Bottom: Corresponding strikelines (perpendicular to the streamlines of the flow) which display an oval-shape flow print similar to the surface flukeprint as seen by the observers. The colors correspond to the modulus of the
velocity at each point.}
\label{fig:t=8}
\end{figure}

To visualize vortex ring formation and the resulting surface patterns, Levy et al. \cite{Levy:2011} conducted a series of experiments in collaboration with Flo-Metrics in San Diego using a tank of
water approximately 122 cm long by 61 cm wide by 49 cm deep.  In water, they oscillated a 2D
fluke shape made of slightly flexible polypropylene (0.16 cm thick) with the boundary
shape scaled up from a photograph of a blue whale fluke.  The mock fluke was 0.2 m tip to
tip, whereas blue whale flukes are typically 5-6 m. The velocity field was computed by particle image velocimetry (PIV) using a laser sheet and small particles similar to Kaleiroscope. The laser sheet illuminating the small flakes was created using a 1 Watt green laser directed through a 3 mm diameter stirring rod.

In each experimental run, oscillations were maintained at 0.32 Hz and no forward motion.
For comparison, blue whale fluke oscillations occur at about 0.5-1.0 Hz with forward
velocity 1-1.5 m/s.  The initial angle of the fluke with the plane of the
surface was varied, and recorded video at 30 fps.  Stills from this video were first reported in \cite{Levy:2011} but did not include PIV.  Here, PIV measurements from several runs initiated at 12.5 and 60 degrees from horizontal are reported.  Each trial consisted of one oscillation from the prescribed angle, down 25 degrees and then up. While the motion of the robotic fluke in the experiments is only a rough approximation of whale motion, the observations are
consistent with the theory described in the previous section. The image processing is completed on a personal computer with the software Davis 7.2 from Lavision. Coordinate $x$ denotes the cruise direction of the whale, $y$ the transverse direction and $z$ the vertical direction.

According to Archer et al. \cite{Archer}, the life-time of a vortex ring interacting with
a free surface at $90^\circ$ of incidence features three stages after creation: approach of the
free surface, slowing and expansion.  The PIV measurements show that an outgoing flow is induced on the surface of water from the center of a patch with an oval shape, see Fig. \ref{fig:t=8}. The induced surface current is anisotropic and time-dependent; it is stronger on the rear of the flukeprint than on the front. Our Fig.  \ref{fig:t=8} (top) is very similar to the numerical simulations reported in \cite{Ohring, Lugt} where a vortex ring impacts obliquely a free surface. On average, the flow is weaker inside the flukeprint than on its boundary. The streamlines (not reported here) features a null-line from which all streamlines radiate outward. The strikelines (perpendicular to the streamlines) display a typical oval shape reminiscent of field observations, see Fig.  \ref{fig:t=8} (bottom).

\begin{figure}[!htbp!]
\includegraphics[scale = .23]{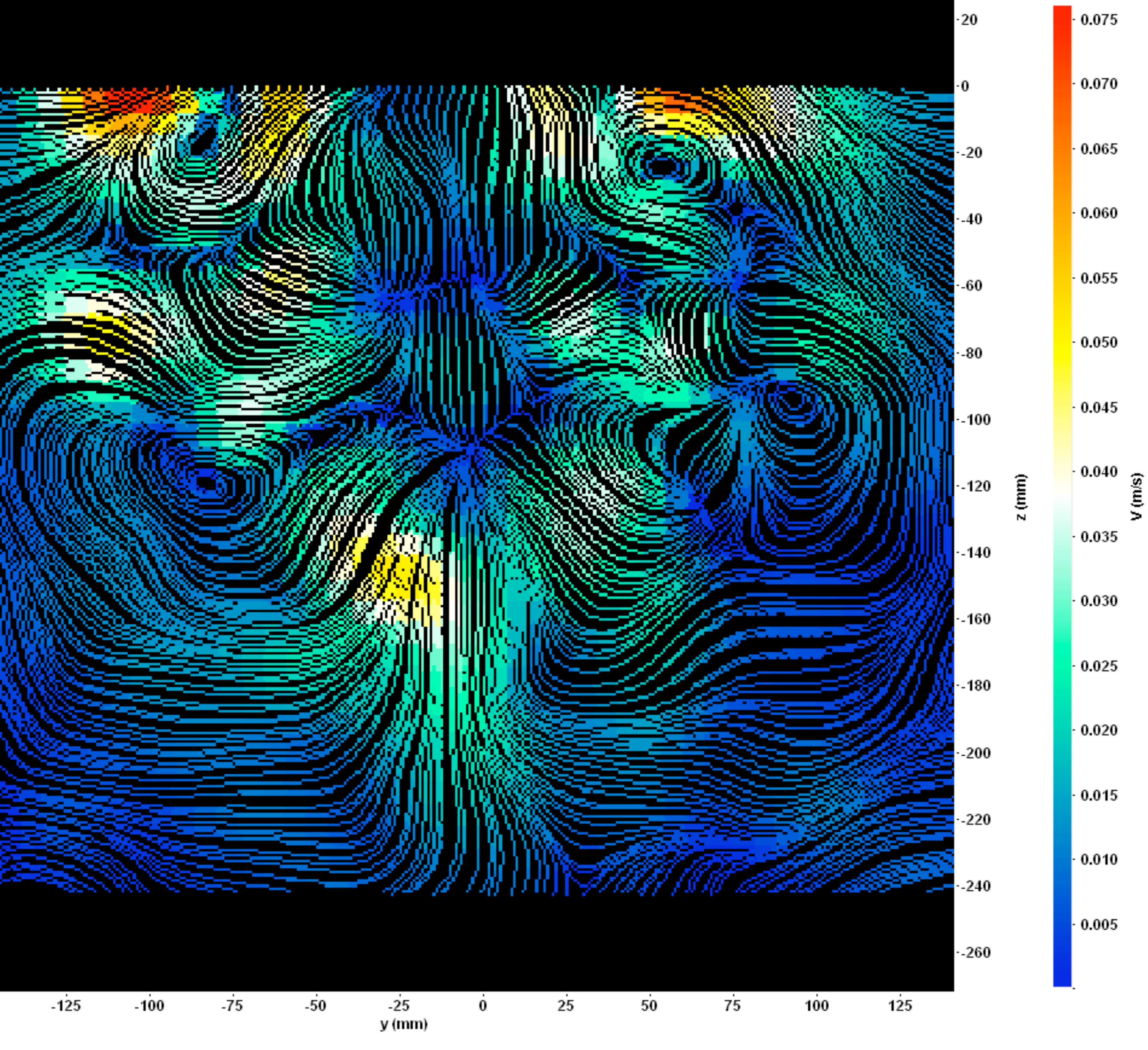}
\includegraphics[scale = .23]{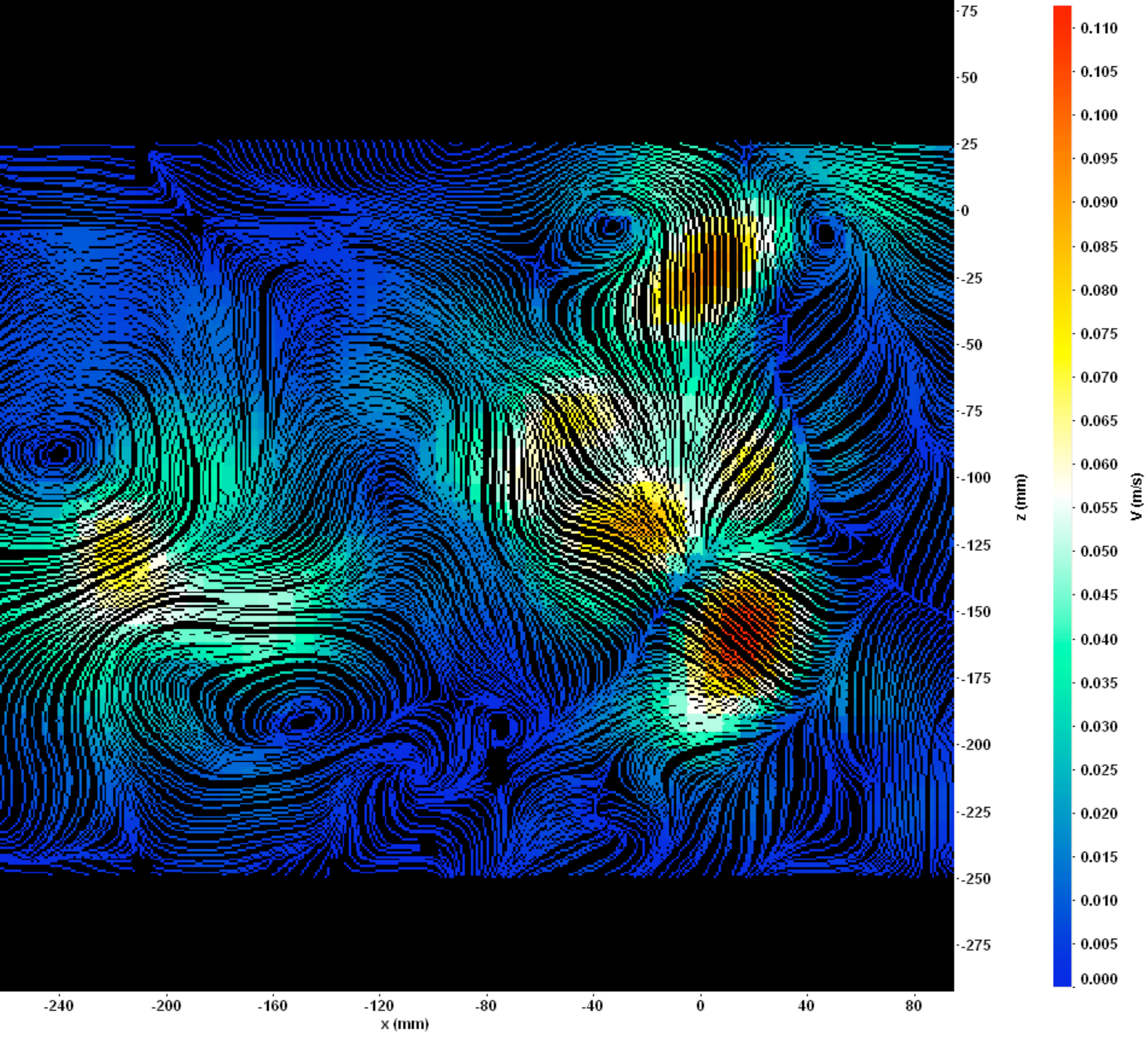}
\caption{Vortex rings. Top: Back view of the whale (t=4.2s after flapping begins). Typical quadrupolar flow on the rear of the fluke. Cross-sections of the two vortex rings. The angle of the fluke is $60^\circ$. Bottom: Side view (t=3.6s after flapping begins). Streamlines of the flow. One observes the cross-section of the two vortex rings created during the downward and the upward motions of the fluke. The upper ring creates free surface shear flows which will block the waves. The colors correspond to the modulus of the velocity at each point. The angle of the fluke is $60^\circ$.}
\label{fig:Views}
\end{figure}

\begin{figure}[htbp!]
\includegraphics[scale = .24]{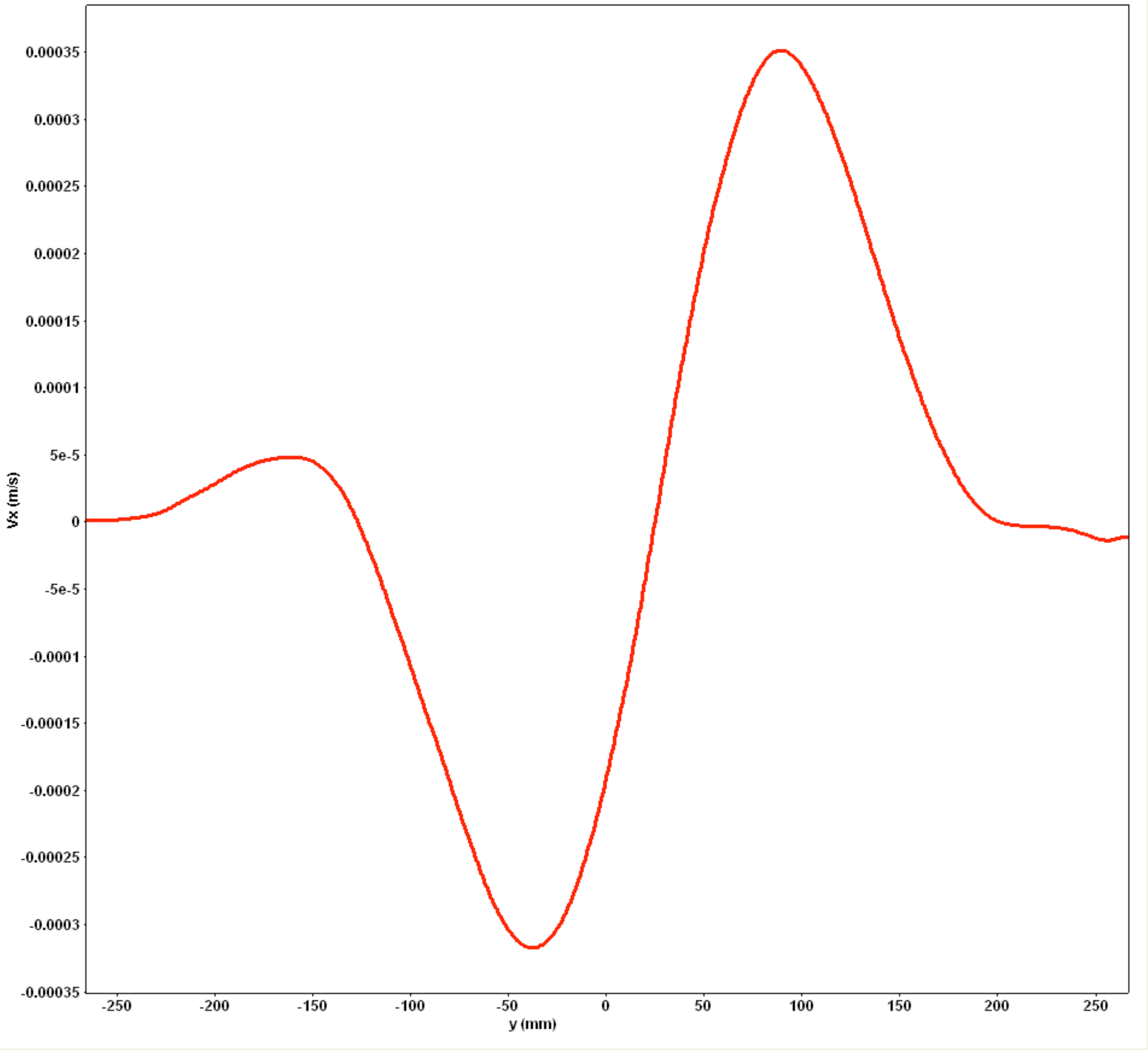}
\includegraphics[scale = .25]{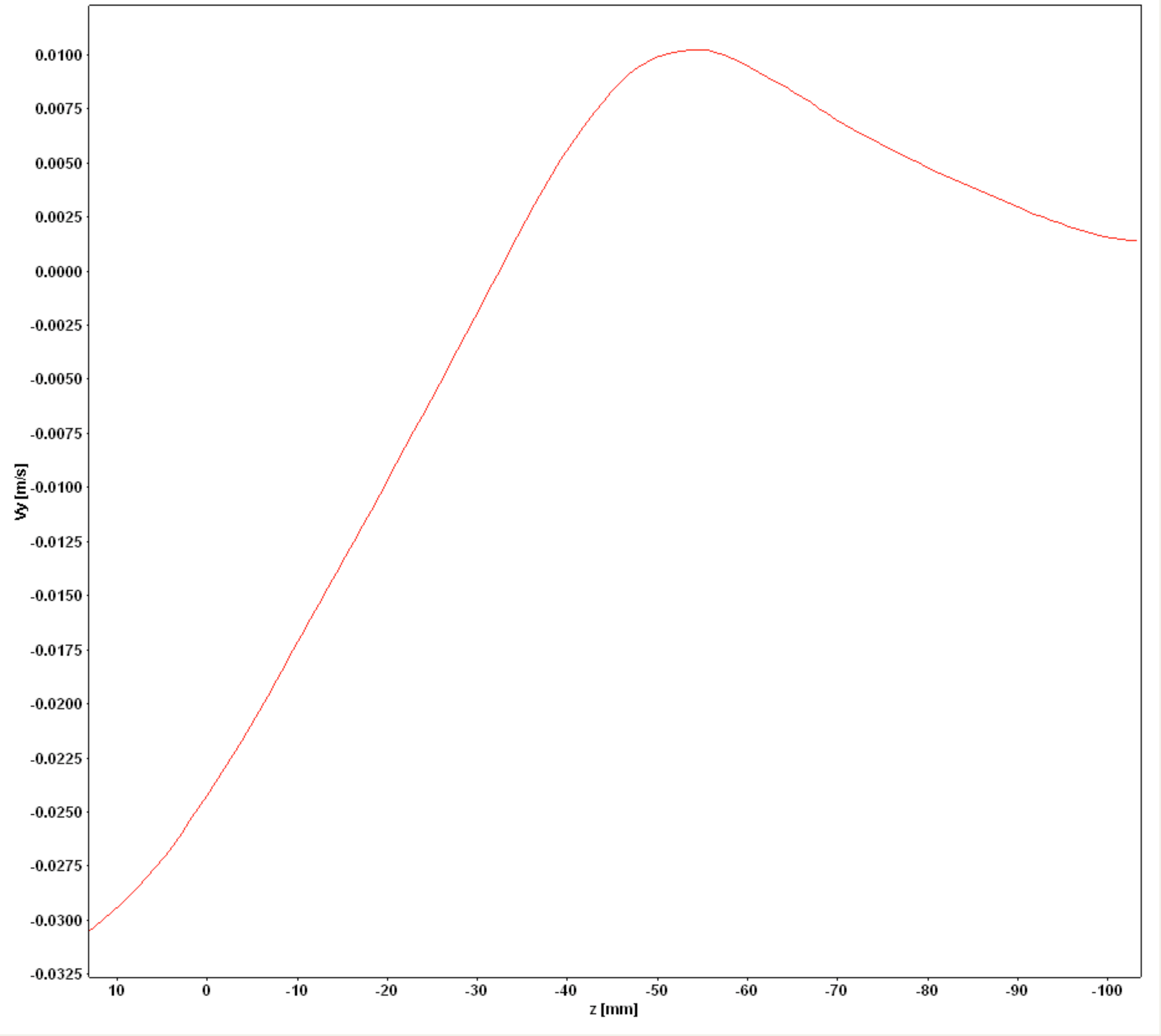}
\caption{Velocity profiles. Top: Back view of the whale (z=-10mm below the free surface). Typical longitudinal variation of the radial surface flow induced by the upper vortex ring (t=2.6s after flapping). Bottom: Back view of the whale (x=-132mm on the left of the upper vortex ring center). Typical vertical variation of the radial surface flow induced by the upper vortex ring (t=8.4s after flapping). Starting at an initial $12.5^\circ$ from the surface, the fluke oscillates through an arc of $\pm25^\circ$.}
\label{fig:PIV}
\end{figure}

In Fig. \ref{fig:Views} (top) a quadrupole is seen behind the mock fluke. The quadrupole corresponds to the successive emissions of two vortex rings observed from the side: one towards the deep in the downward motion of the fluke, one towards the free surface in the upward motion of the fluke, see Fig. \ref{fig:Views} (bottom). Measurements of the surface flow profile in the radial direction demonstrate the inhomogeneity of the flow as one gets far from the center ($y=0$) of the flukeprint, see  Fig. \ref{fig:PIV}. The radial velocity increases almost linearly with the distance to center before reaching a maximum and then decreases to zero.

The radial flow depends on the vertical position as seen on the Fig. \ref{fig:PIV} (bottom).  In these experiments, the PIV indicates that the flow is almost linear with the depth with a flow reversal below the core of the toroidal vortex tube.

As a conclusion, the existing theory for wave-current interactions includes a number of simplifying assumptions.  The first, of course, is that we assume the flow is in one or two dimensions rather than providing a fully 3D model.  Even in the 2D case, our measurements show that the surface flow generated in the whale flukeprint is anisotropic in the radial direction.  Another simplification is the assumption that the flow has usually a uniform velocity profile on the entire depth, or on a given depth \cite{Taylor} or at most a linear velocity profile on the entire depth \cite{Thompson, Biesel, Burns, Tsao, Fenton, Brevik, Kirby, Makarova, Nepf} or on given depth \cite{Taylor}. The PIV displays a flow reversal in the vertical direction in addition to a maximum in the radial direction.

The current approach to treat complicated velocity profiles is to modify the dispersion relation using a depth-averaged flow \cite{Skop, Kirby} but there is no consensus that this result is correctly capturing the physics.  Some of the most promising numerical work in this direction is the research of Archer et al. \cite{Archer} who has computed solutions for flows containing a vortex ring interacting with a free surface: here time dependence can also be taken into account.

The extreme difficulty in describing flukeprint formation by applying our knowledge of wave-current interaction is that we do not have the dispersion relation for the flow induced by the whale. One exception is the work of Skop \cite{Skop}, which considers a simple jet-like flow and assumes a triangular velocity profile in the vertical direction. This jet-like flow could treat the first stage of the flukeprint formation when the vortex ring approaches the free surface. Then, the dispersion relation taking into account a flow reversal during the vortex ring expansion has to be derived theoretically and then solved.\\

The author would like to thank Rachel Levy and David Uminsky for providing the movies of the experiments reported in \cite{Levy:2011}. They also helped the author with many conversations and corrections in the writings. The author thanks David Lee and Justin Holmes for providing the beautiful pictures of the dolphin flukeprints.

\end{document}